\documentstyle[prcrc,11pt,fleqn,epsf]{article}

\title{Determining $\Omega$ from cluster correlation function}

\author{A. Kashlinsky\\NORDITA, Blegdamsvej 17, Copenhagen DK-2100, 
and\\Theoretical Astrophysics Center, Juliane Maries Vej 30, Copenhagen}%
\begin{document}
\maketitle

\begin{abstract} 
It is shown how data on the cluster correlation function can be used 
in order to reconstruct the 
density of the pregalactic density field on the cluster mass scale. The method is 
applied to the data on the cluster correlation amplitude -- richness dependence. The 
spectrum of the recovered density field has the same shape as the density field derived 
from data on the galaxy correlation function which is measured as function of linear 
scales. Matching the two amplitudes relates the mass to the comoving scale it contains 
and thereby leads to a direct determination of $\Omega$. The resultant density 
parameter turns out to be $\Omega$=0.25.
\end{abstract}

\section{Introduction}

This paper presents another way of determining $\Omega$ by 
comparing density fields determined from two independent datasets: APM data
on the galaxy two-point correlation function and the data on the cluster
correlation function -- richness dependence. For reasons that will become clear 
later in the paper, I will reconstruct the quantity which is uniquely related to the
correlation function, $\xi(r)$, of the density field, or its Fourier transform - 
the power spectrum $P(k)$:
\begin{equation}
\Delta^2 = \langle (\delta M/M)^2 \rangle = 4 \pi \int_0^r \xi(r') r'^2 dr'
= 4\pi \int_0^\infty P(k) \left(\frac{3j_1(kr)}{kr}\right)^2 k^2 dk
\end{equation}
I will limit the discussion and the formalism to scales where the present density 
field is linear and therefore can be assumed to reflect the initial conditions, i.e.
$r>r_8\equiv 8h^{-1}$Mpc.

The outline and the idea of the paper are as follows: in Sec.2 I discuss reconstruction
of (1) as function of the linear scale, $r$, from the APM data. In Sec.3 I show how
to use the data on the cluster correlation amplitude -- richness dependence in order
to reconstruct (1) as function of the cluster mass. Comparison of the two density 
fields
gives the amount of mass contained in the given linear scale and provides
a direct determination of $\Omega$. I will show that the two density fields, although
recovered in different and independent ways, give consistent results and require
$\Omega\simeq 0.25$ with a very small uncertainty. For more details the
readers are referred to \cite{k98}.

\section{Density field from galaxy clustering}

First, I establish the rms fluctuation, $\Delta(r)$, as function of the linear scale
$r$ from the APM data on the projected angular correlation function $w(\theta)$. 
For this I use the APM data on $w(\theta)$ \cite{mesl} divided into six narrow 
magnitude 
bins $\Delta m_b$$\simeq$0.5. Galaxies located in each of the bins 
span a narrow(er) range of $z$, so with the data presented in this way  
one can isolate effects of possible galaxy evolution in the entire APM catalog
spanning 17$<$$b_J$$<$20 or 0.07$<$$z$$<$0.2 \cite{k92}. 

The projected angular
correlation function for each bin is related to the 3-dimensional power spectrum of 
galaxy clustering, $P(k)$, via the Limber equation:
\begin{equation}
w(\theta) = \pi \int_0^\infty dz (cdt/dz)^{-1} \phi(z)
\int_0^\infty dk P(k;z) k J_0(\frac{kx(z)\theta}{1+z})
\end{equation}
where $x(z)$ is the comoving distance, $t$ is the cosmic time, and the selection
function is $\phi(z)=(dN/dz/N_{\rm tot})^2$. The latter is related to the range of
magnitudes for galaxies in each bin, $[m_l,m_u]$, and their luminosity function,
$\Phi(L)$, via $dN/dz = (dV/dz) \int_{L(m_u)}^{L(m_l)} \Phi(L;z) dL$ with $V$
being the comoving volume. Eq.(2) allows to relate the data on $w(\theta)$ to the 
underlying power spectrum for each of the six narrow magnitude width bins once all of
the following are specified: the luminosity function in a given band (blue for APM),
the $K$-correction which accounts for the shift along the galactic energy spectrum
resulting from cosmic expansion, and the (possible) galaxy evolution out to the edge of 
the APM sample. For the numbers in the remainder of this section
the luminosity function was adopted from the measurements of \cite{l92};
the $K$-correction was modeled from standard spectra of galaxy populations
\cite{yt88}.

In order to estimate the importance of the (possible) evolution effects, I proceeded
as follows: On very small angular
scales, the data show that the angular correlation function
can be described as a power law, $w(\theta)=A_w \theta^{-\gamma}$ with $\gamma=0.7$. 
On the other hand, as $\theta \rightarrow 0$,
the main contribution in the Limber equation comes from very small linear scales where
the spatial galaxy correlation function can be approximated as 
$\xi(r)$=$(r/r_*)^{-1-\gamma}$ with $r_*=5.5h^{-1}
{\rm Mpc}$. Relating the small-scale $\xi(r)$ to $w(\theta)$ via the Limber 
equation leads to the following expression for $A_w$:
\begin{equation}
A_w = \frac{ \Gamma(\frac{1}{2}) \Gamma(\frac{\gamma}{2}) }
{ \Gamma(\frac{1+\gamma}{2}) } \left(\frac{r_*}{cH_0^{-1}}\right)^{1+\gamma}
 \int_0^\infty \phi(z) \left[\frac{cH_0^{-1}(1+z)}{x(z)}\right]^\gamma
\Psi^2(z) (1+z)^2 \sqrt{1+\Omega z} dz 
\end{equation}
where $\Psi(z)$ accounts for the evolution galaxy clustering, e.g. $\Psi^2(z)
\propto (1+z)^{-3}$ for clustering stable in comoving coordinates. 
Comparing the value of $A_w=w(\theta)\theta^{0.7}$ computed from 
eq.(3) with the data for each bin
allows to constrain the extent to which galaxy evolution 
affects inversion of the APM data in terms of the underlying spectrum. The fits of
the power-law galaxy correlation function to the small-scale data on $w(\theta)$ in
all six narrow magnitude bins show that no galaxy evolution corrections are needed
beyond the normal K-corrections and evolution of the clustering pattern with time
\cite{k98}.

Analysis of the data in narrow magnitude bins without galaxy evolution shows that
the power spectrum obtained by deprojection of the entire APM dataset \cite{be93}
fits the data well in all magnitude bins. Neglecting the dependence on
$\Omega_{\rm baryon}$, CDM power spectra can be parameterized by only two parameters: 
the primordial power index $n$ and the excess power parameter $\Omega h$
\cite{bbks}. Fits of the
CDM models to APM data in narrow magnitude bins at \underline{all} depths show that
the models require $\Omega h$=0.2 if $n$=1, or $\Omega h$=0.3 for tilted models with
$n$=0.7. The latter thus still requires low $\Omega$ in order to fit the
APM data in narrow magnitude bins. {\it CDM models with larger values of
$\Omega h$ would give smaller $w(\theta)$ at large $\theta$; whereas smaller
values would overshoot the data} \cite{k98}.

Left panel in Fig.1 shows the rms fluctuation from the fits to the
APM data vs the comoving 
scale. It is plotted
in units of the fluctuation at $r_8$$\equiv$$8h^{-1}$Mpc, $\Delta(r)/\Delta_8$. 
The reason for plotting
the ratio is that for linear biasing the vertical axis in Fig.1 is independent of the 
bias
factor.

\section{Density field from cluster correlation amplitude -- richness dependence}

In this section I show how, using the data on the cluster correlation amplitude -- 
richness dependence, one can reconstruct the quantity plotted in 
the left panel of Fig.1 as function
of the cluster mass. Comparing the results with $\Delta$ over a given range of $r$ 
gives the amount of mass in the given comoving
scale and leads to a direct determination of $\Omega$.

I assume that clusters of galaxies formed by gravitational clustering \cite{ps84} and 
evaluate their
correlation function based solely on this assumption \cite{k87,k91}. I.e., I assume 
that clusters
of galaxies are identified with regions
that at some early epoch $z_i$ had initial overdensity such that
they would turn-around in less than the age
of the Universe. It is convenient to choose $z_i$ to be sufficiently high when the 
density field on all relevant scales is
in the linear regime. In that case, the amplitude, $\delta_{ta}$, 
of the fluctuation at $z_i$ needed for it to
turn around today at $z$=0 is related to $\Delta_{8,i}$, the amplitude which grows to 
$\Delta_8$=1
at $z$=0, via $\delta_{ta}=Q_{ta}\Delta_{8,i}$. The factor $Q_{ta}$$\simeq$1.65 
for a spherical model and is almost
independent of either $z_i$ or cosmological parameters.

I further assume that the initial density field was Gaussian. In that case the joint 
probability
density to find density contrasts $\delta_{1,2}$ on scales containing masses $M_{1,2}$ 
respectively is given by:
\begin{equation}
p(\delta_1;\delta_2) = 
\frac{1}{(2\pi)^2} \int_{-\infty}^\infty \int_{-\infty}^\infty
\exp(-i{\boldmath\mbox{$q\cdot \delta$}}) 
\exp(-\frac{1}{2} {\boldmath\mbox{$q\cdot C\cdot q$}}) d^2q
\end{equation}
The correlation matrix, $\boldmath\mbox{$C$}$, is related to the spectrum of the 
{\it primordial}
density field: its diagonal elements are the mean square fluctuation $\Delta^2$ on 
scale
containing mass $M$ and the non-diagonal elements are $\simeq$$\xi(r)$ at cluster 
separations $r$ 
greater than the comoving scale subtended by the cluster masses $(\sim$1-3 
$h^{-1}$Mpc).
The probability of two such fluctuations to turn-around by now and thereby form 
clusters of
galaxies is $P_{M_1M_2}$=$\int_{\delta_{ta}}^\infty\int_{\delta_{ta}}^\infty 
p(\delta_1;\delta_2) 
d\delta_1 d\delta_2$. The fraction of such pairs at the present time would be 
$f_{M_1M_2}$=$\partial^2P_{M_1M_2}/\partial M_1/\partial M_2$. The probability for a 
single cluster to form by now
is $P_M = \int_{\delta_{ta}}^\infty p(\delta) d\delta 
$; the 
fraction of such clusters is $f_M$=$\partial P_M/\partial M$.

Now one can construct the correlation function between the present-day clusters of 
different masses.
By definition, the 2-point correlation function of an ensemble of objects with number
density $n$ is given by the probability to find two objects in small volumes $dV_1, 
dV_2$
as $d{\cal P}_{12} = n^2(1+\xi)dV_1dV_2$. 
Since the clusters of mass $M_1,M_2$ will make the fraction $f_{M_1M_2}$ of such pairs,
the probability to find them is
$d {\cal P}_{M_1M_2} = f_{M_1M_2}d{\cal P}_{12}$. On the other hand, the mean number 
density of clusters
of mass $M$ would be $f_M \times n$ and by definition the probability to find two 
clusters is
$d {\cal P}_{M_1M_2} = f_{M_1}f_{M_2} n^2 (1+\xi_{M_1M_2})dV_1dV_2$. Hence the 
correlation function
of clusters of mass $M$ is given by:
\begin{equation}
1+ \xi_{M_1M_2} = \frac{ f_{M_1M_2} }{ f_{M_1}f_{M_2} } (1+\xi_i) 
\end{equation}

Using expansions in terms of Hermite polynomials \cite{js86,k91} leads to the following 
expression \cite{k98}:
\begin{equation}
\xi_{MM}(r) = A_M(\frac{\zeta}{\sqrt{2}}) \xi(r)
\end{equation}
where the amplification factor is:
\begin{equation}
A_M(x) = \sum_{m=0}^\infty \frac{1}{ Q_{ta}^{2m} m!} \xi^m C_m(x)
\end{equation}
with
\begin{equation}
C_m(x) = \frac{ x^{2m} }{4} \left[ \frac{ H_{m+1}^2(x) }{ x^2 } 
+\frac{ H_{m+2}^2(x) }{ (m+1)Q_{ta}^2}\right]
\end{equation}
Here $H_m(x)$ are Hermite polynomials and
$\zeta$=$Q_{ta} \frac{\Delta_8}{\Delta(M)}$, which is directly related to the spectrum of
the primordial density field on scale $M$.  Hence, the data on the cluster correlation 
amplitude -- mass dependence \cite{bs83,bw92} can be 
used to invert eqs.(5)-(7) to directly obtain 
$\Delta(M)$ for the {\it primordial} density field. The accuracy of these
expressions has been confirmed in recent numerical experiments \cite{gs98}.

Since cluster correlation amplitude is known to depend on the cluster richness, ${\cal 
N}$,
one has to translate the richness into cluster mass. I assume that the
two are proportional with the coefficient of proportionality normalized to the data on
the Coma cluster:
\begin{equation} 
M({\cal N}) = M_{\rm Coma} \frac{{\cal N}}{{\cal N}_{\rm Coma}} = 
1.45\times 10^{15} \left(\frac{{\cal N}}{106}\right) h^{-1}M_\odot
\end{equation}
Here, I adopted the values
for the Coma cluster and richness following \cite {kg81}. This assumes that there are no
systematic variations in the galaxy luminosity function in clusters of various masses.

The data on the cluster correlation function can now be used in conjunction with 
this mechanism for cluster formation in order to set further constraints on the cosmological 
models via eqs.(5)-(7). Indeed, the dependence of the cluster correlation function at a given 
richness/mass on $r$ constrains the power spectrum on the scale of cluster separation,
whereas the dependence of the cluster correlation amplitude at a given $r$ on the
cluster richness/mass constrains the shape of the {\it primordial} power spectrum 
on the scale containing that mass. Eqs.(5)-(7) can be used in two ways: 
On the one hand, one can
test cosmological paradigms, such as CDM, by evaluating the cluster correlation
parameters by assuming their power spectrum and comparing computed numbers
for both $\xi_{MM}(r)$ vs $r$ and $A_M$ vs $M$ with observational data. 
Alternatively, one can use these expressions in
conjunction with data on the cluster correlation amplitude in order to directly obtain the
spectrum of the {\it primordial} density field on scale $M$ independently of the assumed
cosmological prejudices. The latter, as I show, can also be used to determine $\Omega$.

In terms of the CDM models, I find that no CDM model, 
whose only free parameters are $(\Omega,h,n)$, can simultaneously fit
three sets of data: the APM data on the galaxy correlation, the data on
the slope of the cluster correlation function with scale at a given
richness, and
the data on the dependence of the cluster correlation amplitude at a given separation
on the cluster richness. Namely, the values of $(\Omega,h,n)$ required by fits to
the APM data ($\Omega h \simeq 0.2$ if $n=1$ and $\Omega h \simeq 0.3$ if $n=0.7$)
would be very different from those required by fits to either of the other two
datasets and vice versa \cite{k98}. 
Accounting for a (slight) dependence on $\Omega_{\rm baryon}$ 
in the CDM transfer function \cite{s95} would not change these conclusions.

In order to invert eq.(6) in terms of $x$, and subsequently $\Delta(M)$,
I used the data on the cluster correlation amplitude -- richness dependence from 
Fig.2 of \cite{bw92}. The amplification coefficient for a given richness/mass
in eqs.(5)-(7) depends on the value of 
$\xi(r)$ on the scale where it is evaluated. The underlying correlation function
$\xi(r)$ enters eq.(6) via the first and higher order terms and can contribute up to 
$\sim$10-30\%
to the total amplification. I chose $r$=$25h^{-1}$Mpc on which to evaluate the amplification
factor from (7). This
scale is sufficiently large compared to $r_8$ to ensure validity of the analysis, 
but where at the same time $\xi(r)$ can be determined sufficiently accurately. In the 
discussion
below I adopted $\xi(25h^{-1}{\rm Mpc})$=0.07 in agreement with the APM data (cf. 
\cite{k98}), but the
numbers that follow
are not very sensitive to varying the value of $\xi(25h^{-1}{\rm Mpc})$ within 
reasonable limits.

\begin{figure}[h]
\centering
\leavevmode
\epsfxsize=1.
\columnwidth
\epsfbox{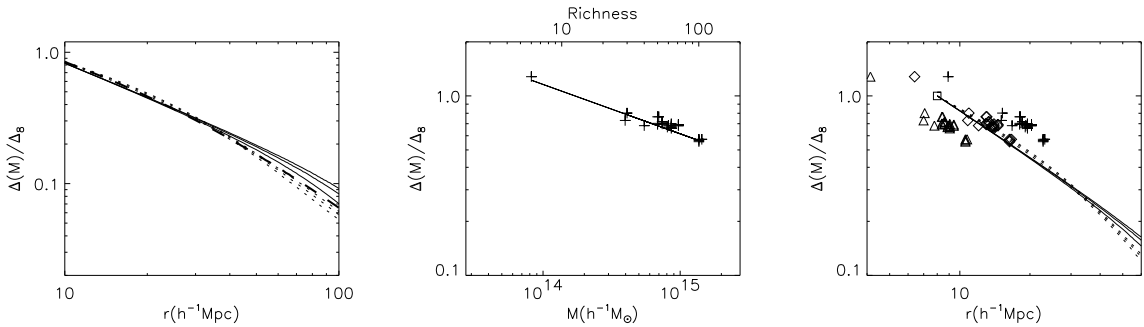}
\caption[]{
(Left)  RMS density fluctuation from APM data is plotted vs 
the comoving scale. It is shown in units of density fluctuation over radius $r_8$.
Solid lines correspond to empirical fit from \cite{k92}; dotted lines to deprojected
spectra from \cite{be93}. Three lines of each type correspond to 1-$\sigma$ 
uncertainty in \cite{be93} and a similar uncertainty in \cite{k92}. Thick dashed line
corresponds to CDM models that fit APM data in narrow magnitude bins: $\Omega h$=0.2 
with $n$=1 or $\Omega h$=0.3 with $n$=0.7. \\
(Middle)  Plus signs correspond to the primordial density field inverted from 
data on cluster correlation amplitude -- richness dependence. Solid line shows the fit
to the points: $\Delta(M)$=$\Delta_8 (M/1.7\times 10^{14}h^{-1}M_\odot)^{-0.275}$.\\
(Right) The values of $\Delta(r)$ from the cluster data for three different values of 
$\Omega$: plus signs correspond to $\Omega$=0.1, triangles to $\Omega$=1
and diamonds correspond to $\Omega$=0.25. Square denotes the (by definition) value of 
unity of $\Delta(M)/\Delta_8$ at 8$h^{-1}$Mpc. The lines represent density fluctuations 
derived from the APM data redrawn from the left panel in the same notation.
}
\end{figure}

Once the data on the rhs of eq.(7) for $A_M$ vs
richness are specified, eqs.(5)-(7) can be solved numerically in order to
obtain $\Delta(M)$.
The middle panel in Fig.1 plots the results of this inversion. The top 
horizontal axis plots the values of ${\cal N}$ at which $\Delta(M)/\Delta_8$ has been 
evaluated. The bottom
horizontal axis shows the mass computed according to the normalization to Coma.
I emphasize again
that this method gives directly the pregalactic spectrum as it was at $z_i$ 
independently of the later
gravitational or other effects. The plot in the middle panel of Fig.1 shows a
clearly defined slope of $\Delta(M) \propto M^{-0.275}$ corresponding to the spectral 
index of
$n \simeq -1.3$. This slope is consistent with the APM implied power spectrum index of 
$w(\theta) \propto \theta^{-0.7}$.

\section{Determining $\Omega$ from the two density fields}

The recovered density field allows one to relate the mass of the cluster to its 
comoving scale
thereby directly determining $\Omega$. The direct fit to the points in the middle panel 
of Fig.1 gives 
\begin{equation}
\Delta(M)$=$\Delta_8 (M/M_8)^{-\alpha}
\end{equation}
with $\alpha$=0.275 and 
$M_8$=$1.7\times 10^{14}h^{-1}M_\odot$. This fit is plotted with solid line in the 
middle
panel. On the other hand, the mass contained in comoving radius $r_8$ is $M(r_8)$=$6.1
\times 10^{14} \Omega h^{-1}M_\odot$. Equating this with $M_8$ leads to $\Omega$=0.28.

Furthermore, one can determine $\Omega$ by comparing $\Delta(r)$ from
the galaxy correlation data over the entire range of the relevant $r$ 
with $\Delta(M)$ derived from the cluster
correlation amplitude -- richness dependence. In order to do this I converted
the numbers for $\Delta(M)$ to those at a given $r$ using that the mass contained
in a given comoving radius in the Universe with density parameter $\Omega$ is
$M(r)$=$1.2\times 10^{12} (r/1h^{-1}{\rm Mpc})^3 \Omega h^{-1}M_\odot$. The right panel 
in Fig.1 shows the values of $\Delta(r)$ from the cluster data for three different 
values of $\Omega$: 0.1 (pluses), 0.25 (diamonds) and 1 (triangles). The square 
denotes the (by definition) value of unity of $\Delta(M)/\Delta_8$ at $8h^{-1}$Mpc. The 
lines represent the APM data redrawn from the left panel. One can see that the spectral 
shape of the density field derived from the cluster data is in good agreement with
that of the APM. {\it The amplitudes of the two fields would match at all $r$ only for 
$\Omega$=0.25}.

\section{Summary and conclusions}

In this presentation I discussed reconstructing density field from the cluster 
correlation amplitude -- richness dependence. It was shown that the data can be inverted 
to obtain the rms density fluctuation in the pregalactic density field on scales 
containing the mass of the clusters. The derived density field has the same spectral 
shape as the density field derived as function of the comoving scale from the APM 
data. Comparing the two amplitudes fixes the amount of mass in given 
comoving scale and allows to determine $\Omega$. The value 
derived from application of the method to the data is $\Omega$=0.25.
This value of $\Omega$, obtained after normalizing the
mass-richness relation to Coma, is in good agreement with that implied by the dynamics 
of the same Coma cluster. 
This further argues that galaxies trace the overall
mass distribution in the Universe.

\end{document}